\makeatletter
\declare@file@substitution{revtex4-1.cls}{revtex4-2.cls}
\makeatother
\documentclass[twocolumn,tighten]{aastex631}
\hypersetup{linkcolor=red,citecolor=blue,filecolor=cyan,urlcolor=magenta}

\usepackage{xcolor}

\shorttitle{AASTeX v6.3.1 Sample article}
\shortauthors{Kharb et al.}
\begin{document}
\title{Relativistic Jet Motion in the Radio-quiet LINER Galaxy KISSR872}
\author[0000-0003-3203-1613]{Preeti Kharb}
\correspondingauthor{Preeti Kharb}
\email{kharb@ncra.tifr.res.in}
\affiliation{National Centre for Radio Astrophysics (NCRA) - Tata Institute of Fundamental Research (TIFR), \\
S. P. Pune University Campus, Ganeshkhind, Pune 411007, Maharashtra, India}
\author[0000-0002-9405-8435]{Eric G. Blackman}
\affiliation{Department of Physics and Astronomy, University of Rochester, Rochester NY, 14627, USA}
\affiliation{Laboratory for Laser Energetics, University of Rochester, Rochester NY, 14623, USA}
\author[0000-0003-0260-3561]{Eric Clausen-Brown}
\affiliation{Microsoft, Greater Seattle Area, Washington, USA}
\author[0000-0001-8996-6474]{Mousumi Das}
\affiliation{Indian Institute of Astrophysics (IIA), II Block, Koramangala, Bangalore 560034, Karnataka, India}
\author[0000-0001-8252-4753]{Daniel A. Schwartz}
\affiliation{Smithsonian Astrophysical Observatory, Cambridge, MA 02138, USA}
\author[0000-0002-0905-7375]{Aneta Siemiginowska}
\affiliation{Smithsonian Astrophysical Observatory, Cambridge, MA 02138, USA}
\author[0000-0002-5331-6098]{Smitha Subramanian}
\affiliation{Indian Institute of Astrophysics (IIA), II Block, Koramangala, Bangalore 560034, Karnataka, India}
\author[0000-0003-3295-6595]{Sravani Vaddi}
\affiliation{Arecibo Observatory, NAIC, HC3 Box 53995, Arecibo, Puerto Rico, PR 00612, USA}

\begin{abstract}
We report superluminal jet motion with an apparent speed of $\beta_\mathrm{app}=1.65\pm0.57$ in the radio-quiet (RQ) low ionisation nuclear emission line region (LINER) galaxy, KISSR872. This result comes from {two} epoch phase-referenced very long baseline interferometry (VLBI) observations at 5 GHz. The detection of bulk relativistic motion in the jet of this extremely radio faint AGN, with a total 1.4~GHz flux density of 5~mJy in the $5.4\arcsec$ resolution Very Large Array (VLA) FIRST survey image and 1.5~mJy in the $\sim5$~milli-arcsec resolution Very Long Baseline Array (VLBA) image, is the first of its kind in a RQ LINER galaxy. The presence of relativistic jets in lower accretion rate objects like KISSR872, with an Eddington ratio of 0.04, reveals that even RQ AGN can harbor relativistic jets, and evidentiates their universality over a wide range of accretion powers. 
\end{abstract}


\section{Introduction} \label{sec:intro}
A distinguishing characteristic of a radio-loud (RL) active galactic nucleus (AGN) is the presence of relativistic jets. Jets in radio-quiet (RQ) AGN when present, are typically small, confined to their host galaxies and slow moving \citep{Kellermann1989, Middelberg04}. Such observational evidence has led to the inference of differing accretion rates and potentially differing accretion disk structures in RL versus RQ AGN. Supermassive black holes (SMBH; M$_\mathrm{BH}\sim10^6-10^9$~M$_\sun$) in RL AGN are believed to be powered at high accretion rates (Eddington ratios, $\mathrm{L_{bol}/L_{Edd}}>0.1$, where $\mathrm{L_{bol}}$ and $\mathrm{L_{Edd}}$ are the bolometric and Eddington luminosities, respectively) which leads to optically thick and geometrically thin disks \citep{Shakura78,Ho2008}. The SMBHs in RQ AGN on the other hand, are powered by lower accretion rate disks with 
$\mathrm{L_{bol}/L_{Edd}}<0.1$ \citep{Ho2008}. These  objects could either have $\mathrm{L_{bol}/L_{Edd}}\sim \mathrm{\dot M}/\mathrm{\dot M_{Edd}}$ in which case they would be low luminosity, high efficiency accretors and also be geometrically thin, or for $\mathrm{L_{bol}/L_{Edd}}\ll \mathrm{\dot M}/\mathrm{\dot M_{Edd}}$ be radiatively inefficient accretion flows \citep[RIAFs;][]{Narayan1998,Narayan+2008,Ho2008} that are geometrically thick but optically thin. Either way, jets with high bulk relativistic speeds such as typically observed in RL AGN have heretofore not been observed in RQ AGN.  

Multi-epoch very long baseline interferometry (VLBI) observations of some relatively radio-bright Seyfert galaxies like NGC1068 and NGC4258, {or radio-quiet quasars like PG1351+640 and Mrk231, have detected sub-relativistic proper motions with typical jet speeds in the range of $0.01c - 0.5c$ \citep{Middelberg2004,Peck2003,Wang2021,Wang2023}.} On the other hand, VLBI observations of radio-loud AGN like blazars \citep[e.g., from the MOJAVE\footnote{Monitoring Of Jets in Active galactic nuclei with VLBA Experiments} sample;][]{Lister2019} reveal apparent jet speeds in the range $0.05c - 35c$. Doppler boosting effects and superluminal motion is a direct consequence of these bulk relativistic jet effects prompting the revised classification of ``jetted'' and ``non-jetted'' AGN in lieu of the radio-loud/radio-quiet divide \citep{Padovani2017}.

In this paper, we present second-epoch VLBI observations of the parsec-scale jet in the radio-quiet type 2 LINER\footnote{Low ionization nuclear emission line region \citep{Heckman1980}} galaxy, KISSR872, which reveals superluminal jet motion. This makes KISSR872 the only radio-quiet (\S~\ref{sec:results}) LINER galaxy so far to exhibit bulk relativistic motion in its jet. Furthermore, this result directly challenges the physical equivalence of jetted/non-jetted AGN classification with the RL/RQ classification.

\subsection{The KISSR Sample: A VLBI Study}
We identified nine nearby (redshifts $z\sim0.03-0.09$) type 2 Seyfert and LINER galaxies from the KISSR\footnote{KPNO Internal Spectroscopic Survey Red} sample of emission-line galaxies \citep{Wegner03} that showed double-peaked or asymmetric narrow emission lines in their SDSS\footnote{Sloan Digital Sky Survey} optical spectra and were detected in the Very Large Array (VLA) FIRST\footnote{Faint Images of the Radio Sky at Twenty-Centimeters} survey with a total flux density $\gtrsim2$~mJy at 1.4~GHz \citep{Kharb2015,Kharb2021}, in order to carry out a VLBI study. We acquired milli-arcsecond (mas) resolution phase-referenced data at 1.5 and 5 GHz with the Very Long Baseline Array (VLBA) from 2013 to 2019 on these sources. These data revealed one-sided radio jets in five of the eight ($>60$\%) VLBA-detected sources \citep{Kharb2021} and dual parsec-scale radio cores in one of them \citep[KISSR102;][]{Kharb2020}. Based on the jet-to-counterjet surface brightness ratios, we obtained limits on the parsec-scale jet speeds assuming jet inclination angles of $\ge50\degr$ as these were type 2 AGN and the half opening angle of the dusty obscuring torus has been suggested to be $\sim50\degr$ \citep{Simpson96}. Type 2 AGN are those with narrow emission lines in their spectra as the broad-lines are expected to be hidden behind an obscuring dusty torus \citep{Antonucci1993}.

This treatment assumes that the jet one-sidedness is a consequence of Doppler boosting/de-boosting effects in approaching/receding jets in these RQ AGN, similar to those in RL AGN. This exercise revealed a range of jet speeds from 0.003c to 0.75c in the KISSR galaxies (larger inclination angles would imply larger intrinsic jet speeds). Interestingly, this speed range is broadly consistent with apparent jet speeds {\it measured} in a handful of radio-bright Seyfert galaxies through multi-epoch VLBI observations \citep{Roy2001, Peck2003, Middelberg2004}. While one-sided jets could also arise due to free-free absorption by the inter-cloud gas in the narrow line regions \citep[NLR;][]{Kharb2021}, this cannot be a favored explanation for jets that are 150~parsec \citep[KISSR434;][]{Kharb2019} or 200~parsec \citep[KISSR872;][]{Kharb2021} long. 

KISSR872 is a type 2 LINER galaxy at the redshift of 0.08255\footnote{from the Sloan Digital Sky Survey Data Release 13 as obtained in January 31, 2017 and listed in the NASA/IPAC Extragalactic Database (NED)}. It has a peak intensity at 1.4~GHz of $4.80\pm0.15$~mJy~beam$^{-1}$ and total flux density of $5.25\pm0.27$~mJy in the VLA FIRST survey ($\theta=5.4\arcsec$).
In this paper, the spectral index $\alpha$ is defined such that flux density at a frequency $\nu$, $S_\nu\propto\nu^\alpha$. The adopted cosmology is H$_0=67.8$~km~s$^{-1}$~Mpc$^{-1}$, $\Omega_\mathrm{mat}=0.308$, $\Omega_\mathrm{vac} = 0.692$, so that at the distance of KISSR872, $1\arcsec$ corresponds to 1.608 kpc.

\begin{figure}
\includegraphics[width=8cm,trim=50 150 50 140]{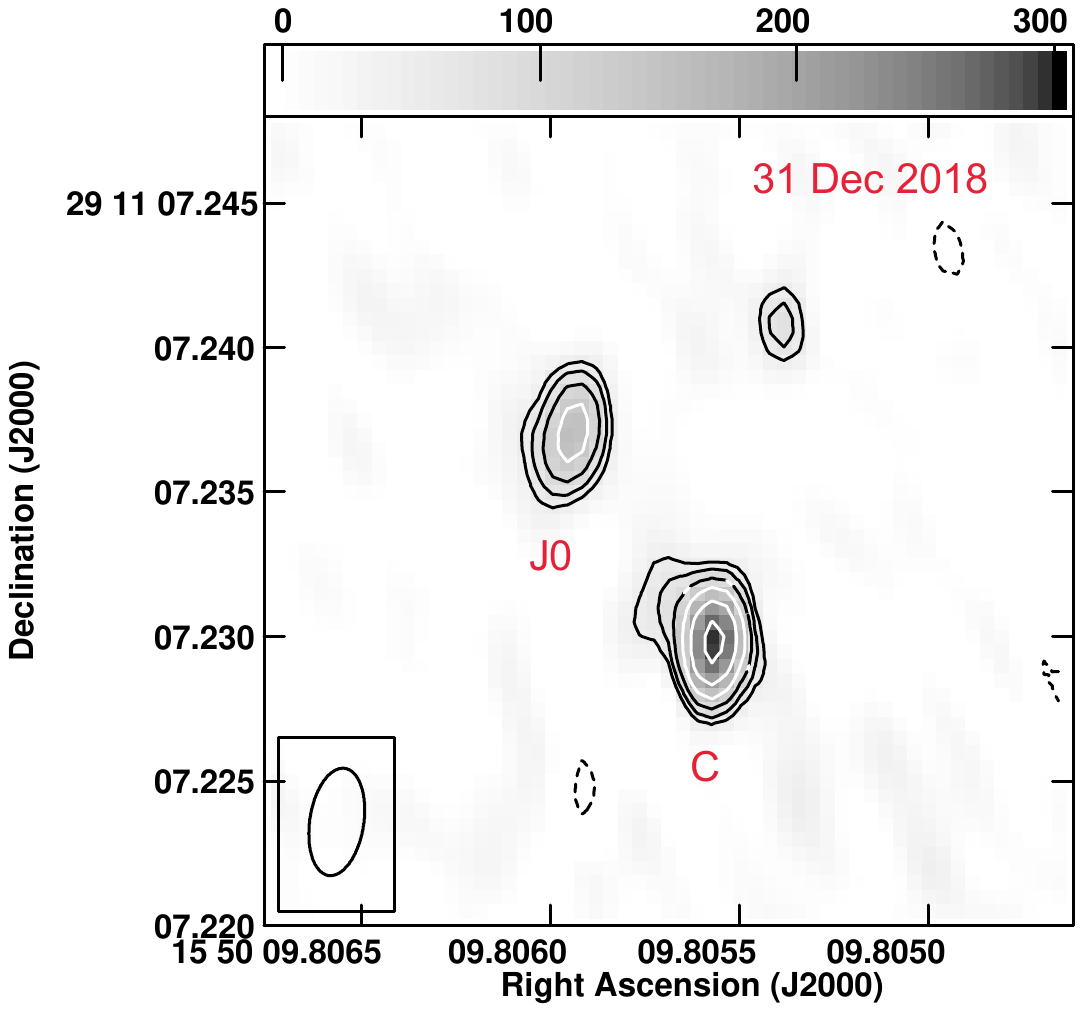}
\includegraphics[width=8cm,trim=50 150 50 140]{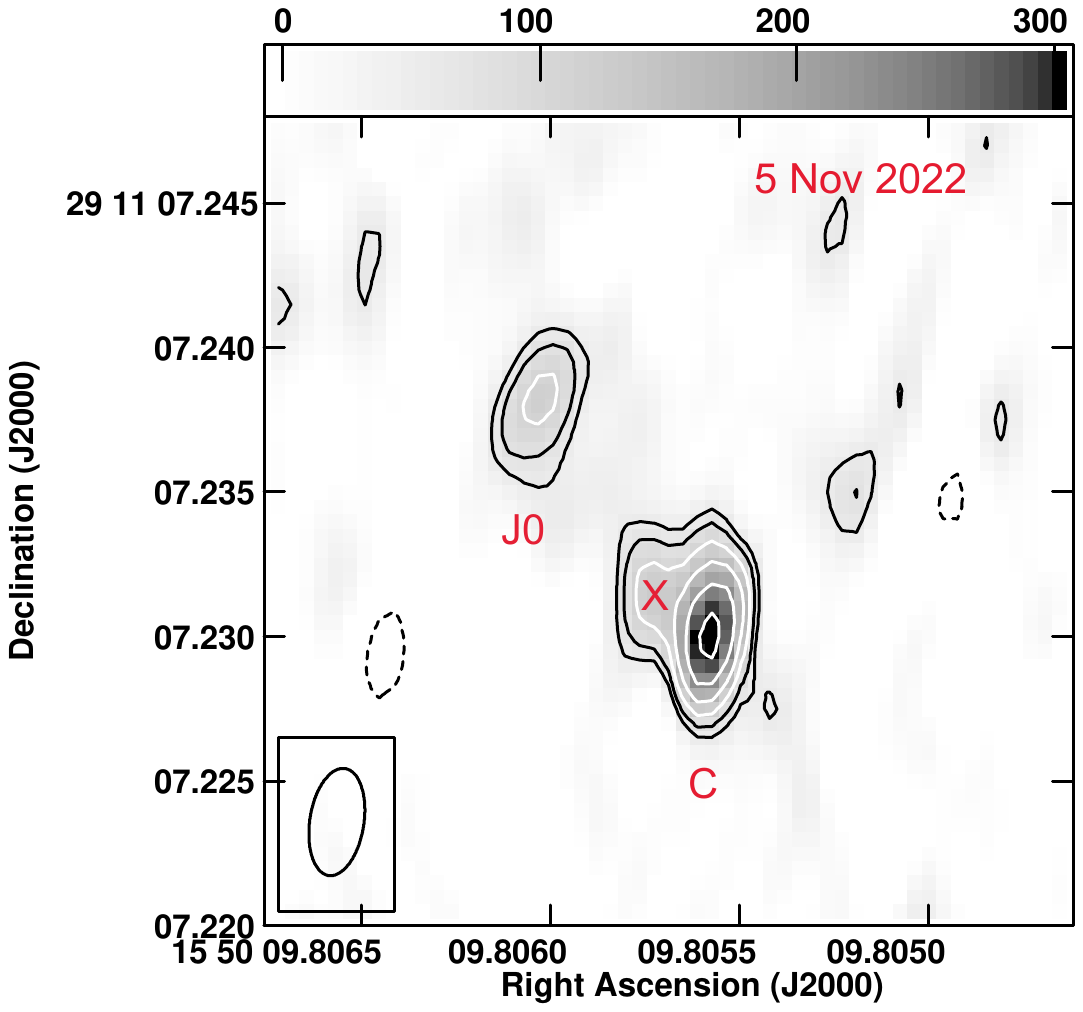}
\caption{\small VLBA image at 5 GHz in contours and greyscale (top) from December 31, 2018, with contour levels $2.962\times (-16, 16, 22.50, 32, 45, 64, 90)~\mu$Jy~beam$^{-1}$, and (bottom) from November 5, 2022, with contour levels $3.410\times (-16, 16, 22.50, 32, 45, 64, 90)~\mu$Jy~beam$^{-1}$. The grey-scale ranges from 1.5 to 301.5 $\mu$Jy~beam$^{-1}$ in both the panels. The beam is $3.75\times1.87$~mas at a PA$=-8.8\degr$ in both figures.} 
\label{fig1}
\end{figure}

\section{Radio Data Analysis}\label{sec:data}
VLBI observations were carried out in a phase-referencing mode using 8 antennas of the VLBA at 1.5~GHz on September 13, 2022 (Project ID: BK246A2); the Kitt Peak (KP) and Pie Town (PT) antennas did not participate in the experiment for technical reasons. Observations at 5~GHz were carried out on November 5, 2022 (Project ID: BK246B4) with 9 antennas of the VLBA; KP did not participate in this experiment due to technical reasons. For both these datasets, 3C345 was used as the fringe-finder while the nearby calibrator 1537+279 was used as the phase-referencing source. The positional error in X and Y directions for 1537+279 was 0.35, 0.81 mas and the separation from KISSR872 was $2.72\degr$. The target and the phase reference calibrator were observed in a ``nodding'' mode in a 5~min cycle (2~min on calibrator and 3~min on source), for good phase calibration.

The data were processed using the VLBA data calibration pipeline procedure {\tt VLBARUN\footnote{URL: http://www.aips.nrao.edu/vlbarun.shtml}} in the Astronomical Image Processing System \citep[{\tt AIPS};][]{Greisen03}. {\tt VLBARUN} is based on the {\tt VLBAUTIL} procedures; their step-by-step usage is described in Appendix C of the {\tt AIPS} Cookbook\footnote{URL: http://www.aips.nrao.edu/cook.html}. The Los Alamos (LA) antenna was used as the reference antenna for both the datasets. The source images were not self-calibrated. All images were created using uniform weighting with {\tt ROBUST +5}. The resultant {\it rms} noise in the images was $\sim 30~\mu$Jy~beam$^{-1}$ at 1.5 GHz and $\sim 20~\mu$Jy~beam$^{-1}$ at 5 GHz. The component positions and flux densities along with errors were obtained with the {\tt AIPS} Gaussian-fitting task {\tt JMFIT} (see Table~\ref{tab1}). {\tt JMFIT} estimates the errors in the Gaussian model fits as a function (inverse square-root) of the component surface brightness, as described by \citet{Condon1997}. We note these errors dominate over the astrometric errors due to the VLBA phase-referencing process, which at the declination of KISSR872 are $\leq50~\mu$arcsec \citep{Pradel2006}. 

\begin{figure}
\includegraphics[width=8.5cm,trim=60 140 40 140]{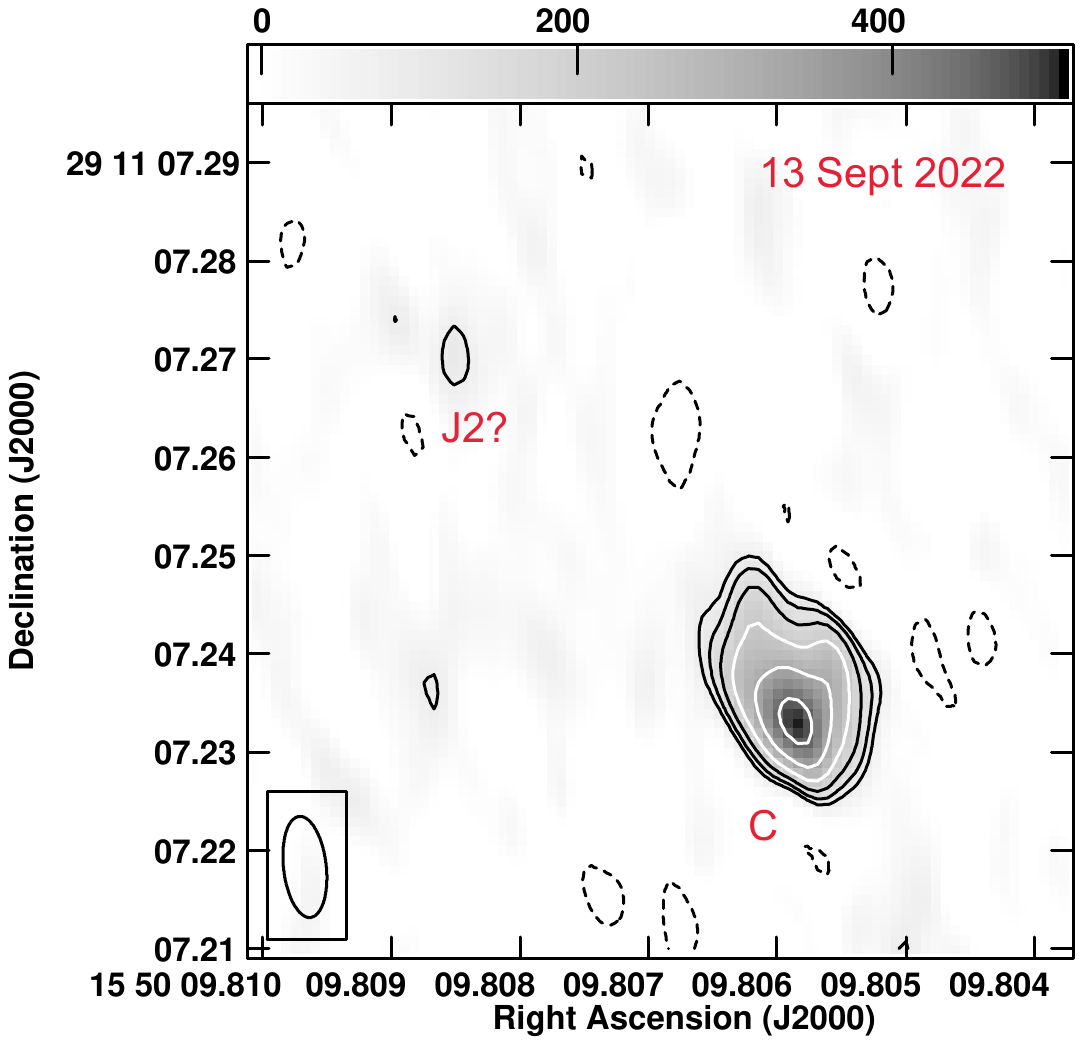}
\caption{\small VLBA image at 1.5 GHz in contours and greyscale from September 13, 2022. The contour levels are $5.055\times(-16, 16, 22.50, 32, 45, 64, 90)~\mu$Jy~beam$^{-1}$. The grey-scale ranges from -2.5 to 508 $\mu$Jy~beam$^{-1}$. Component C is the core-region while component J2 as is noted in the January 2019 data presented in \citet{Kharb2021}. The beam is $10.34\times4.35$~mas at a PA$=6.5\degr$.} 
\label{fig2}
\end{figure}

We created the $1.5-5$~GHz spectral index image for KISSR872 by first matching {the {\tt UV}-spacing} at both frequencies by fixing the {\tt UVTAPER} parameter to 0 to 50 mega$\lambda$, and creating both images with a synthesized beam equal to that of the lower frequency image ($11.01\times4.28$~{mas}, PA=7.3$\degr$). The `core' positions were made coincident using the AIPS task {\tt OGEOM}, before using the AIPS task {\tt COMB} to create the spectral index image. Pixels with flux density values below three times the {\it rms} noise were blanked in {\tt COMB}. A spectral index noise image was created as well. The spectral index and noise values are reported in Section~\ref{sec:results} ahead.

\section{Results}\label{sec:results}
We detect a bright compact component (core or C) and a jet component (J0) in our VLBA 5~GHz image from November 5, 2022 (see Figure~\ref{fig1}), similar to the VLBA 5~GHz image from December 31, 2018. In addition, we see the birth of a new jet component from the core (noted as `X' in Figure~\ref{fig1}). The positions of these components from the two epochs are noted in Table~\ref{tab1}. The identification of component `C' as the core is based both on its relative brightness and its stationarity (with an uncertainity of 90 to 130 micro-arcseconds in the two epochs). It is worth noting here that relatively steep spectrum VLBI `cores', as was observed in the previous epoch for KISSR872, have been observed in other radio-quiet AGN as well \citep[e.g.,][]{Roy00,Orienti10,Bontempi12,Panessa13}. 

The C$-$J0 distance changes from $8.7\pm0.2$~mas ($14.0\pm0.3$ parsec) to $9.8\pm0.4$~mas ($15.8\pm0.6$ parsec) over this time period (time = 121.392 Msec). For a time dilation factor of 1.08255 in the source frame, we measure the apparent speed of the jet component J0 to be $\beta_\mathrm{app}=1.65\pm0.57$. If we assume the bulk flow speed to be $c$, then we can derive a maximum possible viewing angle for the jet components with $\beta_\mathrm{app}=1.65$ as $62\degr$. We also estimate the jet-to-counterjet intensity ratio, R$_J$, in the new 5 GHz image to be $>2.1$ where three times the r.m.s. noise is considered as an upper limit to the counterjet emission. For $\beta=1$, a jet inclination of $<83\degr$ would be implied for a jet structural parameter, $p$, of 3. These inclination angles are consistent with the type 2 classification of KISSR872.

\begin{figure}
\includegraphics[width=8.4cm,trim=50 150 50 140]{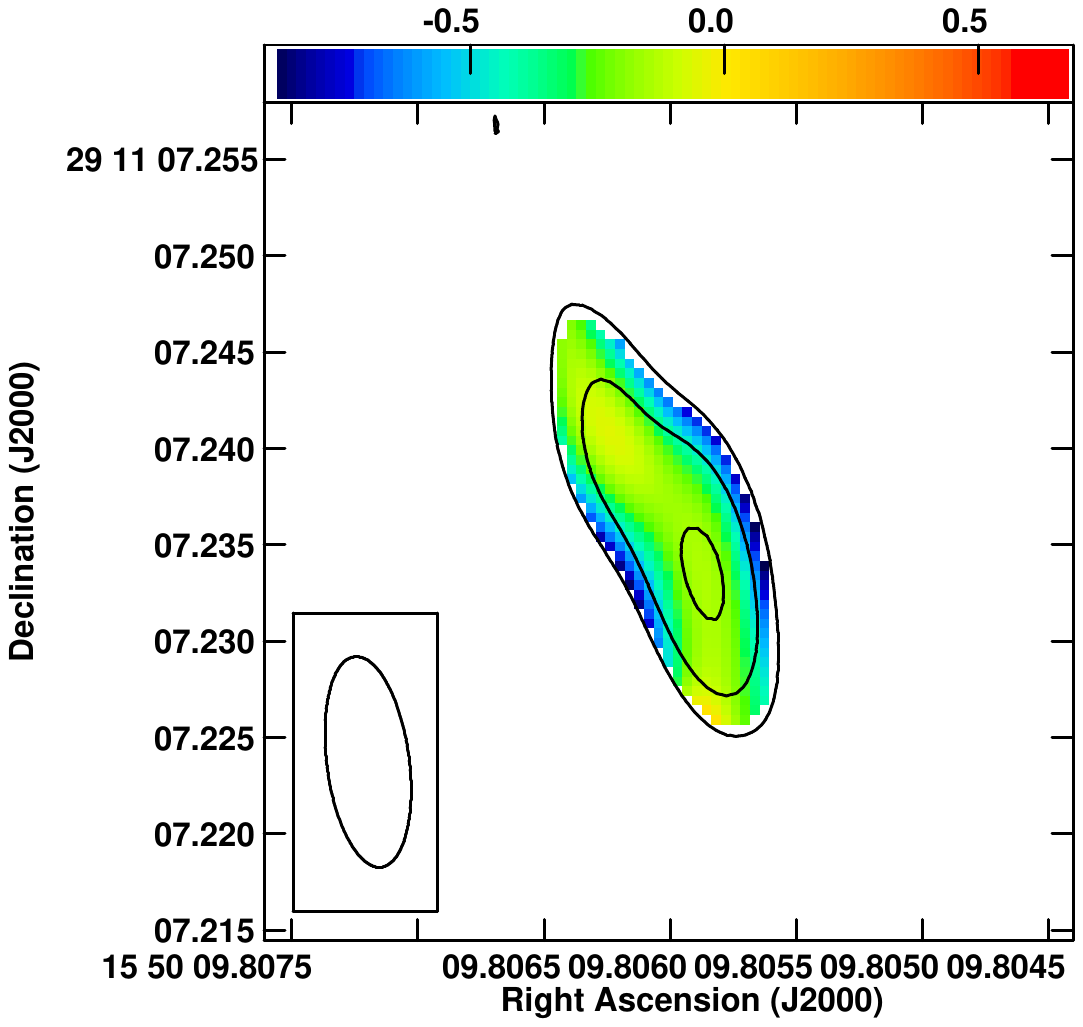}
\caption{\small The $1.5-5$ GHz spectral index image from 2022 in color superimposed by 5 GHz contours at levels $4.77\times(22.5, 45, 90)~\mu$Jy~beam$^{-1}$. The beam is $11.01\times4.28$~mas at a PA= 7.3$\degr$.}
\label{fig3}
\end{figure}

\begin{figure*}
\centering
\includegraphics[width=7.5cm,trim=0 165 0 150]{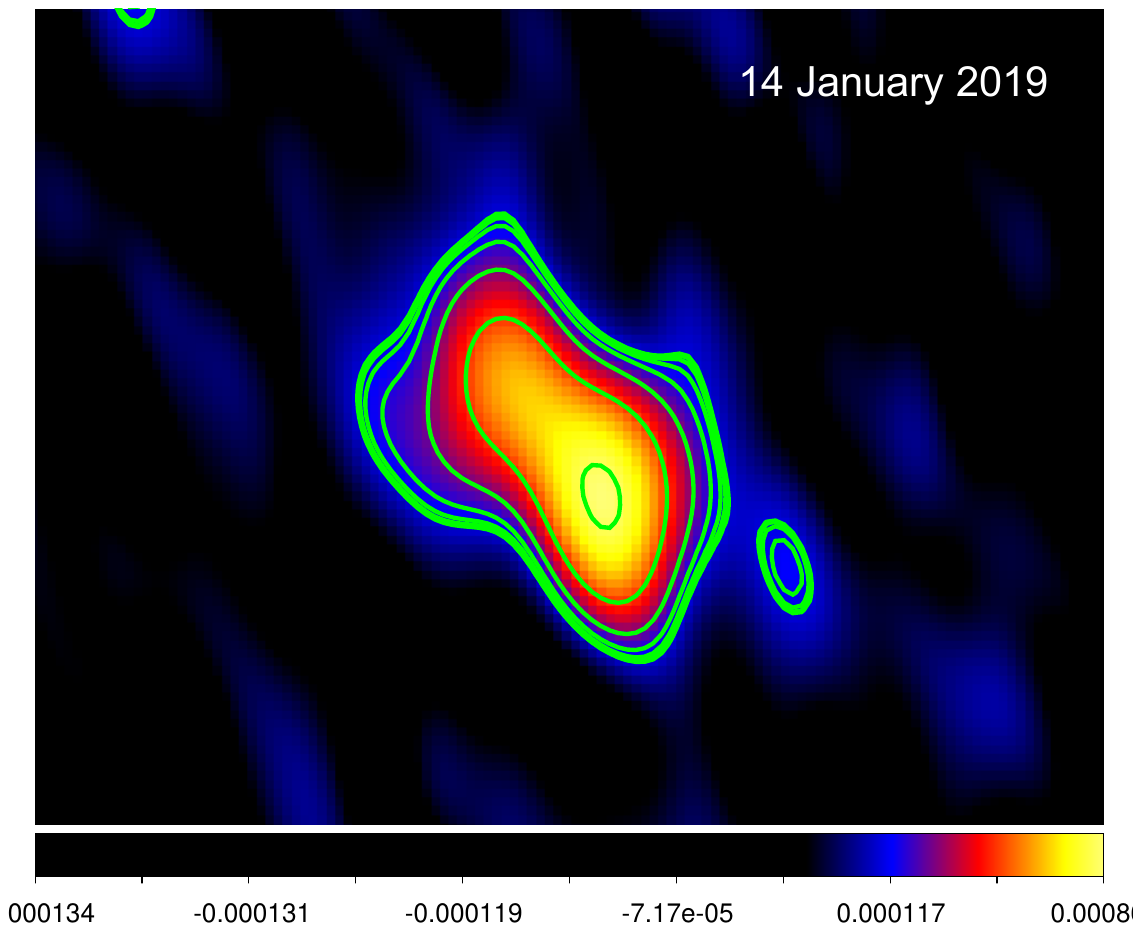}
\includegraphics[width=7.5cm,trim=0 165 0 150]{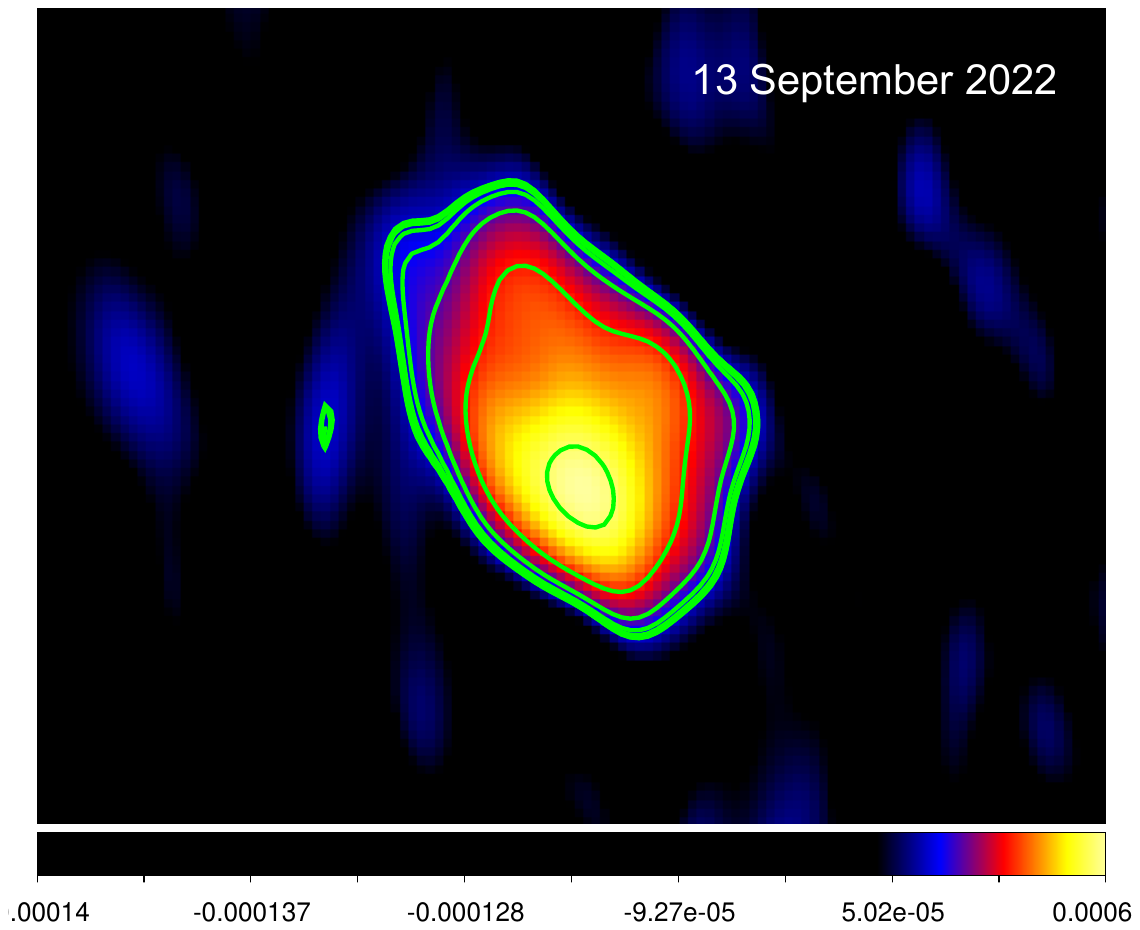}
\includegraphics[width=9.5cm,angle=270]{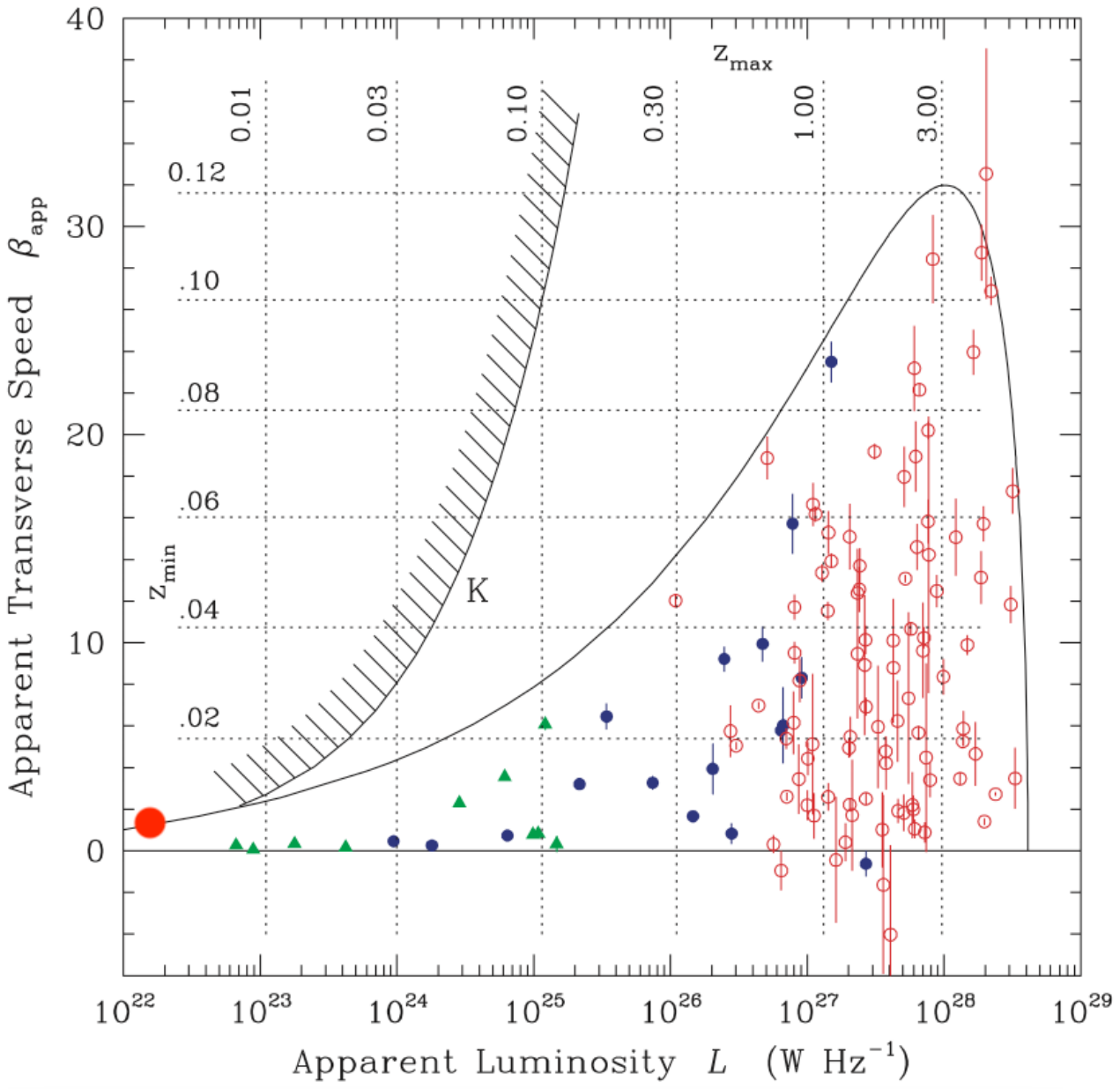}
\caption{\small VLBA image at 1.5~GHz in contours and color with the contour levels being (for top left panel from January 14, 2019) at 95.00, 96.42, 100.89, 115.06, 159.84, 301.47, 749.34~$\mu$Jy~beam$^{-1}$, and (for top right panel from September 13, 2022) at 95.00, 95.87, 98.64, 107.40, 135.09, 222.67, 499.59~$\mu$Jy~beam$^{-1}$. Bottom panel shows the position of KISSR872 as a filled red circle in the speed-luminosity plot of \citet{Cohen2007} for the 2~cm VLBA survey sources.}
\label{fig4}
\end{figure*}

The 1.5~GHz VLBA image from September 2022 shows the `core' structure to have evolved drastically compared to the January 2019 VLBA data (see Figures~\ref{fig3} and \ref{fig4}). The $1.5-5$~GHz spectral index image reveals the average core-jet region spectral index to be $-0.31\pm0.19$ which is also substantially flatter than the average value of $-0.71\pm0.26$ derived in January 2019 \citep{Kharb2021}. One mechanism that can explain the spectral index flattening would be the emergence of a free-free emitting wind component along with the synchrotron jet in the intervening period of four years (see the two panels in Figure~\ref{fig4}). This would give credence to the suggestion of a dynamically changing jet+wind structure being present in Seyfert and LINER outflows \citep{Nevin2018,Kharb2023}. Magnetically collimated jets require a bounding pressure to stabilize and confine the fields \cite[e.g.][]{Begelman+1984,LyndenBell1996}. There do not seem to be any astrophysical jet sources where a bounding envelope or wind is proven to be absent \citep{Blackman+2022}.

In the 1.5~GHz image, we detect a $\sim3.6\sigma$ feature close in position to the jet component J2 as reported in \citet{Kharb2021} (marked as `J2?' in Figure~\ref{fig2}). If this is the same feature as J2, the inferred jet speed would be {unusually large at} $\sim13$c. {Given the faintness of this feature, however, additional data are needed to determine the true nature of this component.} {Nevertheless, on the basis of the high signal-to-noise detection of components C and J0 in the 5 GHz images from two epochs, we conclude that KISSR872 is the first radio-quiet LINER galaxy to show superluminal jet motion in its parsec-scale jet.}

Using the [O III] $\lambda$5007 line luminosity and the relation from \citet{Heckman2004}, the bolometric luminosity (L$_\mathrm{bol}$) in KISSR872 is $2.3\times10^{44}$~erg~s$^{-1}$. A black hole mass of $4.4\times10^7$~M$_{\sun}$ has been estimated in KISSR872 using the stellar velocity dispersion ($\sigma$) from spectral line fitting and the M$_\mathrm{BH} - \sigma$ relation from \citet{McConnell2013}. The Eddington luminosity ($\equiv 1.25\times10^{38}$~M$_\mathrm{BH}$/M$_{\sun}$) is $5.6\times10^{45}$~erg~s$^{-1}$ and the Eddington ratio is 0.04. Its radio-to-optical luminosity ratio, R = L~$(\mathrm{5~GHz})$/L~$(\mathrm{4858~\AA}) = 1.59\times10^{22}/6.17\times10^{21}~\mathrm{[W/Hz]} = 2.6$, places it firmly in the radio-quiet AGN category. Here we have used nuclear AGN properties \citep[see][]{Ho2008} by using the VLBI 5~GHz luminosity and the SDSS g-band `PSF' luminosity from NED; g-band being the closest in wavelength to the B-band used in the original work of \citet{Kellermann1989}. Finally, the jet power, Q$_\mathrm{jet}$ of $2.6\times10^{43}$~erg~s$^{-1}$ is estimated using the 5~GHz radio luminosity, L$_\mathrm{R} = 8.0\times10^{38}$~erg~s$^{-1}$, and the relativistic-beaming-corrected relations of \citet{MerloniHeinz2007}. Interestingly, this value lies in the range of Q$_\mathrm{jet}$ values estimated for radio-loud FRI radio galaxies \citep{Rawlings91}.

\begin{table*}
{\scriptsize
\caption{VLBA component properties in 5~GHz images}
\begin{center}
\begin{tabular}{cclcccccc}
\hline\hline
{Date} & {Comp} & {Position (h~m~s, $\degr~\arcmin~\arcsec$)}& {{$\theta_\mathrm{maj}$ (mas), $\theta_\mathrm{min}$ (mas), PA ($\degr$)}}& {I$_\mathrm{peak}$ ($\mu$Jy~beam$^{-1}$)} & {S$_\mathrm{total}$ ($\mu$Jy)} \\\hline
05-NOV-2022 & C & RA 15 50 09.805578 $\pm$ 0.000004 &4.7 $\pm$ 0.3, 2.1 $\pm$ 0.1, 174 $\pm$ 3& 334.09 $\pm$ 22.1 & 471.93 $\pm$ 48.6 \\
            &   & DEC 29 11 07.2301 $\pm$ 0.0001  && & \\
            &X  & RA 15 50 09.805743 $\pm$ 0.00001 & 4.2 $\pm$ 0.7, 2.3 $\pm$ 0.4, 169 $\pm$ 11& 123.88 $\pm$ 22.2 & 167.76 $\pm$ 47.4 \\
            &   & DEC 29 11 07.2316 $\pm$ 0.0003  && & \\
            &J0 & RA 15 50 09.806027 $\pm$ 0.00001 &5.2 $\pm$ 0.9, 2.7 $\pm$ 0.5, 161 $\pm$ 10& 121.23 $\pm$ 21.5 & 239.93 $\pm$ 60.4 \\
            &   & DEC 29 11 07.2380 $\pm$ 0.0003  && & \\
31-DEC-2018 & C & RA 15 50 09.805569 $\pm$ 0.000003 &3.6 $\pm$ 0.2, 2.0 $\pm$ 0.1, 179 $\pm$ 4& 299.01 $\pm$ 17.2 & 303.90 $\pm$ 30.2 \\
            &   & DEC 29 11 07.22982 $\pm$ 0.00008  && & \\
            &J0 & RA 15 50 09.805944 $\pm$ 0.000007 &3.9 $\pm$ 0.4, 2.1 $\pm$ 0.2, 170 $\pm$ 6& 160.34 $\pm$ 17.1 & 192.23 $\pm$ 33.5 \\
            &   & DEC 29 11 07.2370 $\pm$ 0.0001 && & \\
\hline
\end{tabular}
\end{center}
{\small Column 1 notes the dates of the 5~GHz VLBA observations. Column 2 indicates the VLBA components in the 5~GHz images as shown in Figure~\ref{fig1}. Column 3 notes the component positions in right ascension and declination as provided by the Gaussian-fitting AIPS task JMFIT. {Column 4 provides the major and minor axes of the fitted Gaussian components in milli-arcseconds along with their position angles in degrees.} Column 5 notes the peak surface brightness while Column 6 lists the total flux density at 5~GHz. }
\label{tab1}}
\end{table*}

\section{Discussion}
A jet speed of $\sim1.25-2.66$c has been detected in the Seyfert 2 galaxy, IIIZw2 \citep{Brunthaler00}. However, IIIZw2 is a radio-intermediate AGN with $\sim50$ kpc ($\sim30$~kpc on one side) jets \citep{Silpa2021a}. The production of jets in AGN likely requires magnetically mediated extraction of energy and angular momentum either from a fast rotating SMBH via the \citet{Blandford77} mechanism, a magnetic tower from field lines linking the accretion disk and the black hole \citep{LyndenBell1996}, a jet anchored in the surrounding accretion disk \citep{Blandford82}, or some combination of the two that evolves in time \cite[e.g][]{Sikora+2013}. The magnetic jet must also itself be stabilized by a surrounding pressure. The presence of a magnetically arrested disk \citep[MAD;][]{Narayan2003} with outflows driven by the \citet{Blandford77} mechanism, has been invoked to explain collimated and highly magnetised jets in RL AGN. While this may dominate for RL sources, the VLBI core-shift study of \citet{Chamani2021} report that the magnetic field strength limits derived for IIIZw2, with the caveat that ``equipartition'' between magnetic and particle energy is assumed, are too small for a MAD disk to be operational. This does not, however, preclude a magnetically mediated jet. 

{The radio-quiet quasar Mrk231 typically exhibits sub-relativistic jet speeds in the range of $0.013-0.14c$ \citep{Ulvestad1999,Wang2021} but can turn superluminal ($v>3.15c$) during flaring states \citep{Reynolds2017}. Similarly, a three epoch VLBI monitoring study of the radio-quiet quasar PG1351+640 by \citet{Wang2023} shows a sudden increase in its jet speed over the recent two epochs, even though its overall speed remains sub-relativistic at $0.37c$. Such a behaviour (of varying speeds for individual jet knots over time) is also observed in the long-term VLBI monitoring studies of radio-loud blazars \citep[e.g.,][]{Cohen2007}, underscoring the importance of continuous VLBI monitoring of AGN jets to full understand their nature. \citet{Sbarrato2021} have highlighted the case of relativstic jets in two high redshift ($z>5$) RQ quasars, viz., SDSS J0100+2802 and SDSS J0306+1853, based on spectral energy distribution (SED) modelling, and suggested that relativistic jets may be more common in the early Universe.}

The case of KISSR872 exemplifies that relativistic jets can persist in lower accretion rate engines. Their jets may  benefit from  having phases of a thick disk mode of accretion \citep{Begelman+1982,Sikora+2013,LyndenBell2003}, as would be the case with a RIAF, although we cannot know if this source is a RIAF without an independent estimate of the accretion rate. A thick disk can help supply large scale flux and may also help to collimate a jet from the central black hole, but a broad wind pressure confining a magnetic jet from a thin disk is also possible. The maximum four-velocity of the jet in typical models is determined, by the product of the angular speed at which the foot points are anchored times the radius out to which field lines remain quasi-rigid, i.e., the Alfv\'en radius \citep{Blandford1979,Blandford82,Pelletier+1992,LyndenBell2003}. That can be relativistic even for a modest magnetic field if the black hole is spinning rapidly. The extent to which the disk versus the black hole is the predominant anchoring rotator for a jet's magnetic field in intermediate and weak radio sources is not well constrained, as a faster spinning black hole would also mean a closer inner-most stable orbit of the disk. Only for jets so powerful that they exceed the available accretion power is extraction from the black hole essential, but this is not the case for KISSR872. 

{It is worth noting that using optical/UV (o) and X-ray (x) data on LINER galaxies, \citet{Maoz2007} concluded that there may be  no sharp changes in the source SEDs at the lowest luminosities; thin AGN accretion disks could persist at low accretion rates. Using the \citet{Maoz2007} LINER sample, \citet{Sobolewska2011} showed that the $\alpha_{ox}$ values of LINERs were consistent with their accretion disks being in a hard spectral state, a state which produces jets, analogous to the case of Galactic X-ray binaries. }

Interestingly, KISSR872 falls {close to the `aspect curve' generated by \citet{Cohen2007} for an intrinsic Lorentz factor of 32 and an intrinsic luminosity of $10^{25}$~W~Hz$^{-1}$, which envelopes the apparent transverse speed and (15~GHz) radio luminosity data for radio-loud AGN belonging to the 2~cm VLBA survey \citep{Kellermann1998}. These data support the idea that the `relativistic beaming model' works for the entire population of 2~cm VLBA sources, and perhaps also KISSR872. In the case of KISSR872, the 15~GHz apparent radio luminosity ($L=1.2\times10^{22}$~W~Hz$^{-1}$) was estimated using the VLBA 5~GHz total flux density assuming an $\alpha=-0.3$}. This result highlights the universality of jets over a wide range of accretion and radio powers and raises questions about the presence of intrinsically different central engines in RL and RQ AGN. The reduced spread in superluminal speeds of the low luminosity sources would be consistent with lower Lorentz factors of the jet at the loci of observation, and thus a need to be more favorably inclined close to the inclination angle to the observer that maximizes the apparent jet speed.

\section{Summary \& Conclusions}
We report the detection of superluminal jet motion and thereby the presence of a relativistic jet in the radio-quiet LINER galaxy KISSR872. Unlike the other Seyfert galaxy showing superluminal jet motion, viz., IIIZw2, which has $\sim50$~kpc jets and is largely a radio-intermediate AGN, any kpc-scale outflow in KISSR872 is $\leq8.7$~kpc in extent, that being the resolution of the VLA FIRST image ($\theta=5.4\arcsec$). Moreover, only 5 mJy of the radio flux density is present within this 8.7~kpc region at 1.4 GHz, 44\% of which is present in the 200~pc-scale region probed by the VLBA \citep[see][]{Kharb2021}, making this an extremely radio-quiet AGN. The results from our VLBI study of KISSR872 demonstrate that its engine can produce relativistic jets that can influence the NLR clouds giving rise to double-peaked emission lines in its optical spectra. The presence of relativistic jets in an extremely radio-quiet AGN like KISSR872 also challenges any suggestion of intrinsically different central engines in RQ AGN compared to RL AGN and highlights the universality of jets over a wide range of accretion powers. The position of KISSR872 in the ($\beta_{app}, L$) plot of radio-loud AGN by \citet{Cohen2007} is consistent with these suggestions. The conclusion for black hole engines resonates with similar conclusions reached for disk-jet sources from non-relativistic engines like protostars \citep{Narang2023}. It remains to be determined what engine geometry, and thus the exact mechanism of jet collimation, is operating in a low luminosity AGN like KISSR872.

\vspace{5mm}
\facilities{VLBA, SDSS}

\bibliography{ms}{}
\bibliographystyle{aasjournal}

\section*{Acknowledgements}
{We thank the referee for a careful reading of our manuscript and providing us with suggestions that have improved the manuscript significantly.}
We thank Zsolt Paragi for their insightful suggestions that have improved this manuscript. PK acknowledges the support of the Department of Atomic Energy, Government of India, under the project 12-R\&D-TFR-5.02-0700. PK acknowledges the support of the Chandra X-ray Center, and the Center for Astrophysics, Harvard \& Smithsonian, as a visiting scientist. MD acknowledges the support of the Science and Engineering Research 
Board (SERB) grant CRG/2022/004531 for this research. AS and DAS are supported by the NASA Contract NAS8-03060 to the Chandra X-ray Center of the Smithsonian Astrophysical Observatory. EB acknowledges support from National Science Foundation  grant PHY-2020249.
The National Radio Astronomy Observatory is a facility of the National Science Foundation operated under cooperative agreement by Associated Universities, Inc. Funding for the Sloan Digital Sky Survey has been provided by the Alfred P. Sloan Foundation, the Heising-Simons Foundation, the National Science Foundation, and the Participating Institutions. SDSS acknowledges support and resources from the Center for High-Performance Computing at the University of Utah. The SDSS web site is www.sdss.org. This research has made use of the NASA/IPAC Extragalactic Database (NED), which is operated by the Jet Propulsion Laboratory, California Institute of Technology, under contract with the National Aeronautics and Space Administration. 
\end{document}